\documentclass[10pt,a4paper]{article}
\usepackage{hyperref}
\pdfoutput=1
\usepackage{color}
\usepackage{comment}
\usepackage{amsmath}
\usepackage{pgf}
\usepackage{tikz}
\usetikzlibrary{arrows,automata}
\usepackage{sgame}
\usepackage{float}
\usepackage{authblk}

\begin{document}

\title{Evolutionary Exploration of the Finitely Repeated Prisoners' Dilemma--The Effect of Out-of-Equilibrium Play\footnote{Also appears as: \emph{Lindgren, K.; Verendel, V. Evolutionary Exploration of the Finitely Repeated Prisoners’ Dilemma--The Effect of Out-of-Equilibrium Play. Games 2013, 4, 1-20.}}}
\author[1]{Kristian Lindgren\thanks{kristian.lindgren@chalmers.se}}
\author[1]{Vilhelm Verendel\thanks{vive@chalmers.se}}
\affil[1]{\small Complex Systems Group, Department of Energy and Environment, \hspace{2cm} Chalmers University of Technology, G\"oteborg, Sweden}


\maketitle

\begin{abstract}\noindent The finitely repeated Prisoners' Dilemma is a good illustration of the discrepancy between the strategic behaviour suggested by a game-theoretic analysis and the behaviour often observed among human players, where cooperation is maintained through most of the game. A game-theoretic reasoning based on backward induction eliminates strategies step by step until defection from the first round is the only remaining choice, reflecting the Nash equilibrium of the game. We investigate the Nash equilibrium solution for two different sets of strategies in an evolutionary context, using replicator-mutation dynamics. The first set consists of conditional cooperators, up to a certain round, while the second set in addition to these contains two strategy types that react differently on the first round action: The "Convincer" strategies insist with two rounds of initial cooperation, trying to establish more cooperative play in the game, while the "Follower" strategies, although being first round defectors, have the capability to respond to an invite in the first round. For both of these strategy sets, iterated elimination of strategies shows that the only Nash equilibria are given by defection from the first round. We show that the evolutionary dynamics of the first set is always characterised by a stable fixed point, corresponding to the Nash equilibrium, if the mutation rate is sufficiently small (but still positive). The second strategy set is numerically investigated, and we find that there are regions of parameter space where fixed points become unstable and the dynamics exhibits cycles of different strategy compositions. The results indicate that, even in the limit of very small mutation rate, the replicator-mutation dynamics does not necessarily bring the system with Convincers and Followers to the fixed point corresponding to the Nash equilibrium of the game. We also perform a detailed analysis of how the evolutionary behaviour depends on payoffs, game length, and mutation rate.
\end{abstract}

\section{Introduction} 
\label{sectionIntroduction}

During the past two decades there has been a huge expansion in the development and use of agent-based models for a variety of societal systems and economic phenomena, ranging from markets of various types and societal activities such as energy systems and land use, see e.g., \cite{ComputationalEconomicsHandbook2,ArtificialStockMarket1997,ABMFinancial2007,WITCH2006,SimLUC2003} for a few illustrative examples. A key issue in the construction of such models is the design of the agents. To what extent are the agents rational? And what does it mean that an agent is rational? This has divided scholars into different camps. Herbert Simon \cite{Simon1957Bounded} introduced the concept of "bounded rationality", which can be implemented in a variety of ways in agent-based models, assuming that there is some limitation on the reasoning capability of the agent. Game theory provides useful methods for the analysis of interaction between agents and their behaviour. But it is also well known, from experiments of human behaviour in game-theoretic situations \cite{CamererBehavioral}, that human subjects do not always follow the behaviour predicted by game theory -- and for good reasons. People can in many cases establish cooperation for mutual benefit where game theory would predict the opposite. This discrepancy between "rational" game-theoretic agents and human ones is often attributed to either limited rationality of human reasoning or to social preferences. The latter can in some cases be referred to as rule-based rationality \cite{Aumann2008RuleRationality} under which rules-of-thumb may have developed over time in cultural evolution under positive selective feedback from the benefits of cooperation. 

In the modeling and construction of agents it is therefore of high importance that the assumptions made on rationality and the reasoning process are made explicit. Binmore discusses this in his classic papers "Modeling rational players" \cite{Binmore1987Modeling,Binmore1987Modeling2}. He makes the distinction between {\em eductive} and {\em evolutive} processes leading to an equilibrium in a game. The former refers to a process internal to the agent representing a reasoning process, while the latter may work with much simpler characterisation of the agents where evolutionary processes lead to an equilibrium by mutation and selection on the population level. In evolutionary game theory and agent-based modeling it is common to use a combination of these, but one seldom designs agents who carefully reason about possible actions and their consequences. And the fundamental question still remains: What does it mean for an agent to be rational? 

One of the major achievements in game theory is the establishment of the Nash equilibrium concept and the existence proof that any finite game has at least one such equilibrium \cite{Nash1950}. The Nash equilibrium is a situation where no player can gain by unilateral change of strategy, and in that sense this can be seen as a rational equilibrium in providing the player a best response to other players' actions. The Nash equilibrium is thus often regarded as a result of rational reasoning, reflecting the behaviour of rational players. Importantly, for our discussion, this view of the Nash equilibrium has also been carried over from single-shot to repeated games.

In several finitely repeated games, in which the number of rounds is known, the solution of how players choose actions can be guided by the backward induction procedure. This is often exemplified by the Prisoners' Dilemma, for which the single round game has a unique Nash equilibrium with both players defecting, while the indefinitely repeated game has an uncountable infinity of equilibria allowing for cooperation. However, when the exact number of rounds, $n$, is known, a player can start with considering the last round, in which the score is maximized by defecting. So, with both players being rational in the sense that they want to maximize their own score, the outcome of the last round is clear---mutual defection. But then the next-to-last round turns into the last unresolved round, and the same reasoning applies again resulting in mutual defection also for round $n-1$. The assumption needed is that each player knows that the other one is rational. The procedure then repeats all the way to the first round, showing that the Nash equilibrium is mutual defection from the start of the game. Since the result of the backward induction procedure does not seem to lead to an intuitively rational result it is often referred to as the "backward induction paradox" \cite{PetitSugden1989Backward}. Note that backward induction also applies to certain one-shot games, and the discrepancy between theory and observation has been discussed also for several such examples, e.g., the "Beauty contest" \cite{BoschDomenech2002} and the "Traveler's dilemma" \cite{Kaushik1994}.

There are at least two important objections against the generality of the reasoning based on backward induction. The first objection is empirical, since studies on how human players behave in the game show a substantial level of cooperation, but with a transition to lower levels of cooperation towards the end. Explanations are several, and this implies that several mechanisms are in play. For example, it has been observed in the laboratory that subjects cooperate initially but attempt to cheat each other by deviation in the end \cite{SeltenStocker1986,CamererBehavioral}.

The second objection is conceptual and strongly connected to the notion of rationality and what can be considered as a rational way of reasoning. The only equilibrium that can exist in a given finite repetition is the Nash equilibrium, but whether that is to be considered as rational is the question. A critical point concerns what conclusion a player should draw if the opponent deviates from what backward induction implies and instead cooperates \emph{in the first round}. It is then clear that the opponent is not playing according to the Nash equilibrium, and there is a chance to get a higher score for some time if cooperation can be established.

In this situation, the choice between (i) following backward induction and defecting from start and (ii) deviating from backward induction by starting with cooperation becomes a strategic decision. One can imagine "rational" players in both categories. In the first category, there are then two options, either one just plays defect throughout the game whatever the opponent does, as backward induction suggests, or one switches to cooperation if the opponent cooperates. In the second category, both players are again faced with the question of, provided the opponent is cooperating, when to switch to defection. Obviously, there cannot exist a fixed procedure for deciding on when it is optimal to switch from cooperation to defection, since such a strategy would be dominated by the one that switches one round before. However, the interaction and survival of different ways to handle first round cooperation can be studied using evolutionary methods.

The purpose of this paper is to investigate in an evolutionary context the performance of strategies representing the strategic choices discussed above in the finitely repeated Prisoners' Dilemma. In Binmore's terms, we focus on an evolutive process, in which each agent has a certain, relatively simple strategy for the game, and the mix of strategies and their evolution is investigated on the population level. Importantly, the chosen strategies can all be seen as components in the reasoning processes discussed above: both (i) the steps involved in the backward induction process, and (ii) the steps initiating and responding to cooperation in the first round which then reflects the possibility for strategies to deviate from equilibrium play. It is well known that evolutionary drift or mutations, at least if sufficiently strong, can drive the population away from a fixed point corresponding to the Nash equilibrium. Under what circumstances does the evolutionary dynamics lead to the same result as the backward induction process with a Nash equilibrium as its fixed point, and when can deviation from Nash equilibrium play alter that process? The answer, which is elaborated in this paper, depends on choices of a number of critical model characteristics and parameters: selected strategy space, mutation rate, payoff matrix, and the length of the game.

We prove that, for a simple set of strategies, {\em i.e.}, conditional cooperators, the replicator-mutation dynamics is always characterised by a stable fixed point corresponding to the Nash equilibrium in the limit of zero (but positive) mutation rate. Numerically we show that such a result does not hold in general, even if the only Nash equilibrium is characterised by defection from the start of the game. The strategy set that allows for different reactions on the first round may lead to a path of actions different from what is considered in the backward induction process. On the population level, this turns out to destabilise the dynamics, and for a large part of parameter space, the evolution does not bring the system to a stable fixed point, even in the limit of zero (but positive) mutation rate. The dynamics is instead characterised by oscillatory behaviour.

Most of the work related to evolutionary dynamics, backward induction, and the finitely repeated Prisoners' Dilemma concerns the replicator dynamics \cite{SchusterSigmund1983RepliatorDynamics} with no mutation, as this class has been examined analytically more thoroughly than the replicator-mutation dynamics \cite{HofbauerSelectionmutation1985}. Nachbar \cite{Nachbar1992Evolution} studies convergence in the dynamics and shows that for 2-stage games all cooperation goes extinct when starting from a mixed population. This result is extended by Cressman \cite{Cressman1996Evolutionary, cressman2003evolutionary} to the finitely repeated Prisoners' Dilemma of arbitrary length.

Several authors have investigated various types of evolutionary dynamics under the effect of perturbations or mutations (\cite{Noldeke1993, CressmanSchlag1998Dynamic, BinmoreSamuelsson1999Drift, Sergiu2002, HofbauerSandholm2011}). Here the focus has primarily been on other games than the finitely repeated Prisoners' Dilemma. Gintis et al. \cite{GintisCressmanRuijgrok2009} consider the replicator dynamics subject to recurrent mutation when mutation rate goes to zero. For finite noncooperative games, they show that the dynamics need not converge to the subgame perfect equilibrium, and limiting equilibria can be far away from the this equilibrium. They also show that in the $n$-player Centipede game, there exists a unique limiting equilibrium as mutation rate goes to zero, which is far from the subgame perfect equilibrium but equal in payoffs. 

Ponti \cite{Ponti2000Cycles} studies replicator-mutation dynamics in the Centipede game \cite{Rosenthal1981Centipede}. Using simulations, he finds recurrent phases of cooperation in the evolutionary dynamics, and for particular payoffs of the game, this is shown to depend both on mutation rate as well as the length of the game. It is left as an open question whether such behaviour would disappear in the long run, or persist for negligible mutation rate. This result is most relevant to the present paper in the context of the finitely repeated Prisoners' Dilemma.

Our work is also related to the literature on the "backward induction paradox" \cite{PetitSugden1989Backward}, which has focused on deviation from equilibrium play in the first round of extensive games. For the backward induction reasoning to be rational, it has been assumed that each player has full knowledge about the rationality of the other player, and that both players know this, et cetera, known as "common knowledge of rationality". It has been proved by Aumann \cite{Aumann1995CKR} that such common knowledge implies backward induction in games with a unique subgame perfect equilibrium. However, backward induction provides no firm basis to act when a player deviates from equilibrium play \cite{Aumann1996Reply,Binmore1997Backward,Gintis2010Towards}. Our study can be viewed as an evolutionary study of a population in a setting where reacting to out-of-equilibrium play in the first round is possible.

\section{Evolutionary Dynamics}
\label{sect_evoldynamics}

The evolution of strategies in the finitely repeated Prisoners' Dilemma is studied using replicator dynamics with a uniform mutation rate. This is a model of an infinite population where all interact with all, and in which each strategy $i$ occupies a certain fraction $x_i$ of the population. The selection process gradually changes the population structure, based on a comparison of the average score $s_i$ of strategy $i$ with the average score $s$ in the population, 
\begin{align}
s_i &= \sum_{j=1}^n u(i, j) x_{j} \\
s &= \sum_{i=1}^n s_i x_i \;
\end{align}
where $u(i,j)$ is the score for strategy $i$ meeting strategy $j$ in the $N$-round Prisoners' Dilemma. Assuming a uniform mutation rate of $\epsilon$, the replicator-mutation dynamics can be written,
\begin{eqnarray}
\frac{dx_i}{dt} = x_i \left( s_i - s - \epsilon \right) + \epsilon/n \;
\label{eqnMutatorReplicator}
\end{eqnarray}
where $n$ is the number of different strategies. The dynamics conserve the normalisation of $x$.

The score $u(i,j)$ between two strategies, $i$ and $j$, is calculated from $N$ rounds of the Prisoners' Dilemma with a payoff matrix for the row player given by
\begin{center}
\begin{game}{2}{2}
& $C$ & $D$\\
$C$ &$1$ &$0$\\
$D$ &$T$ &$P$
\end{game} \\
\end{center}
This means that the reward $R$ for mutual cooperation (C, C) has been set to 1, and the "suckers payoff" $S$ for cooperation against defection (C, D) is 0. The temptation to defect against a cooperator (D, C) is associated with a score $T$, and for mutual defection (D, D) the score is $P$. With the usual constraints of $T > R > P > S$ and $2R > T + S$ it remains to study the parameter region given by $T \in [1,2]$ and $P \in [0,1]$. There are in principle three independent parameters in the game, but in combination with the replicator-mutation dynamics, the third one is incorporated in the mutation rate $\epsilon$ (which can be derived by subtraction and division by $S$ and $R-S$, respectively, in the replicator-mutation dynamics equation, Equation~(\ref{eqnMutatorReplicator})).

\newpage
\subsection{Selecting the Strategy Set}
\label{subsect_selectstrat}

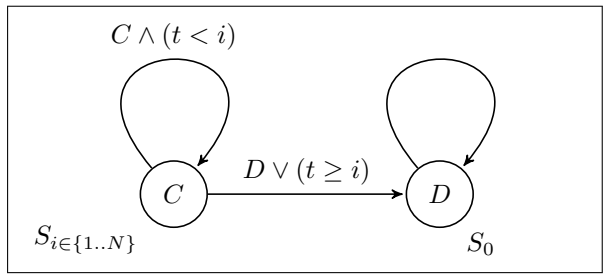
\begin{figure}
\center
\framebox{
\begin{tikzpicture}[->,>=stealth',shorten >=1pt,auto,node distance=3.5cm,semithick]
\tikzstyle{every state}=[circle,draw]

\node[state, label=220:$S_{i \in \{1..N\}}$] (C) {$C$};
\node[state, label=300:$S_0$] (D) [right of=C] {$D$};

\path (C) edge node {$D \lor (t \geq i)$} (D) (C) edge [loop above, distance=2.5cm, out=130, in=50, looseness=0.8] node {$C \land (t < i)$} (C) (D) edge [loop above, distance=2.5cm, out=130, in=50, looseness=0.8] node {} (D);
\end{tikzpicture}
}
\caption{Finite state machine illustrating the first strategy set, $\Gamma_1$, of the $S_i$ strategies. The action to perform is in the node, and transitions are taken on the basis of the opponent's action in the previous round (or based on the previous round number $t$). All strategies start in the left node (C), except $S_0$ that starts with defection in the right node (D). $S_i$ cooperates conditionally for $i$ rounds after which it starts to defect.}
\label{fig_statespace0}
\end{figure}

We investigate the evolutionary behaviour considering two sets of strategies. The first one is a strategy set $\Gamma_1$ that represents various levels of depth in applying the backward induction procedure to conditional cooperation. A strategy in this set is denoted $S_k$ (with $k \in \{0, 1, ..., N\}$), which means that the strategy is playing conditional cooperation up to a round $k$ and then defects throughout the game, see Figure~\ref{fig_statespace0}. Conditional cooperation means that if the opponent defects, one switches to unconditional defection for the rest of the game. For example, $S_N$ is prepared to cooperate through all rounds (or as long the opponent does), while $S_0$ defects from the first round. It is then clear that strategy $S_k$ dominates $S_{k+1}$ (for $k \in \{0, 1, ..., N-1\}$), and backward induction leads us to the single Nash Equilibrium in which both players choose strategy $S_0$.

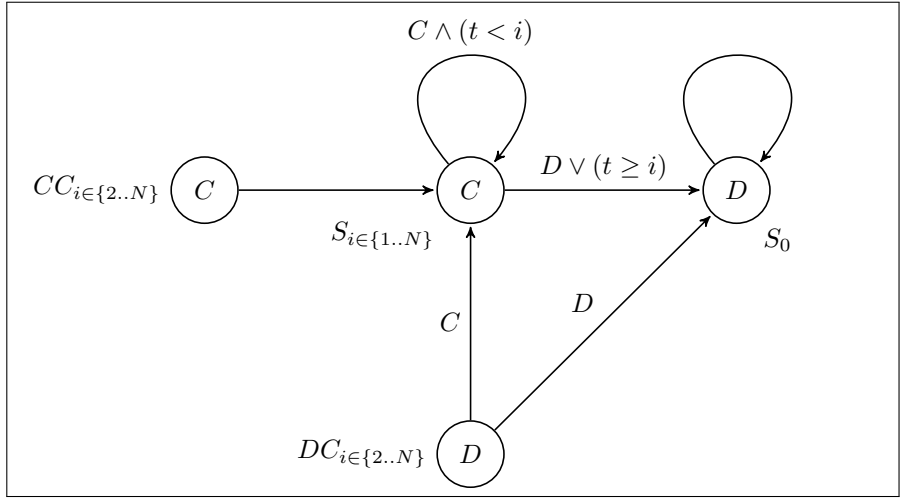
\begin{figure}
\center
\framebox{
\begin{tikzpicture}[->,>=stealth',shorten >=1pt,auto,node distance=3.5cm,semithick]
\tikzstyle{every state}=[circle,draw]

\node[state, label=220:$S_{i \in \{1..N\}}$] (C) {$C$};
\node[state, label=left:$CC_{i \in \{2..N\}}$] (A) [left of=C] {$C$};
\node[state, label=left:$DC_{i \in \{2..N\}}$] (B) [below of=C] {$D$};
\node[state, label=300:$S_0$] (D) [right of=C] {$D$};

\path (A) edge node {} (C) (B) edge node {$C$} (C) (B) edge node {$D$} (D) (C) edge node {$D \lor (t \geq i)$} (D) (C) edge [loop above, distance=2.5cm, out=130, in=50, looseness=0.8] node {$C \land (t < i)$} (C) (D) edge [loop above, distance=2.5cm, out=130, in=50, looseness=0.8] node {} (D);
\end{tikzpicture}
}
\caption{\label{fig_statespace1} Finite state machine illustrating the extended strategy set $\Gamma_2$ consisting of the strategies $S_i,CC_i$, and $DC_i$. $S_i$ are the conditional cooperators as described in Figure~\ref{fig_statespace0}. The Convincers are denoted $CC_i$ and the Followers $DC_i$. Strategies start in the state at which the name is placed. The strategies $CC_i$ start with at least two rounds of cooperation, which may trigger $DC_i$ to switch from defection to cooperation. After that the latter two strategies act as the conditional cooperators $S_i$.}
\end{figure}

Note that the entire strategy set $\Omega$ for the $N$-round Prisoners' Dilemma is very large, as a strategy requires a specification how to react on each possible history (involving the opponent's actions) for every round of the game. This results in $2^{2^N-1}$ possible strategies, e.g. for $N=10$ rounds this is in the order of $10^{308}$. The selection of strategies to consider is critical. One can certainly introduce strategies so that other Nash equilibria are introduced along with those characterized by always defection. For example, selecting only three strategies, e.g., $\{S_0,S_5,S_{10}\}$ will lead to a game with three equilibria, one for each strategy. But this is an artefact of the specific selection made. Here, we do not want to create new Nash equilibria but we want to investigate how and if the evolutionary dynamics brings the population to a fixed point dominated by defect actions corresponding to the original Nash equilibrium. One way to achieve this is to make sure that iterated elimination of weakly dominated strategies can be applied to the constructed strategy set, in a way so that only strategies that defect throughout the game remain, keeping the non-cooperative characteristic of the Nash equilibrium.

The second set of strategies $\Gamma_2$ (in Figure~\ref{fig_statespace1}) that we consider is given by extending $\Gamma_1$ to include also strategies corresponding to steps of reasoning in which one (i) tries to establish cooperation even if the opponent defects in the first round and (ii) responds to such attempts by switching to cooperation for a certain number of rounds. We refer to such strategies as "Convincers" and "Followers", respectively. 

A Convincer strategy $CC_k$ starts with cooperation twice in a row, regardless of the opponent's first round action, and in this way it can be seen as an attempt to establish cooperation even with first round defectors. From round 3, the strategy plays as $S_k$, {\em i.e.}, conditional cooperation up to a certain round $k \in \{2, 3, ..., N\}$.

A Follower strategy $DC_k$ starts with defection, but can be triggered to cooperation by a Convincer (or $S_k$ where $k>0$). So the Follower switches to cooperation in the second round, after which it, like the Convincer, plays as $S_k$, {\em i.e.}, conditional cooperation up to round $k \in \{2, 3, ..., N\}$. (A Follower strategy is also triggered to cooperation by an $S_k$ strategy, but $S_k$ does not forgive the first round defect action and cooperation cannot be established.) 

For the extended strategy set, $\Gamma_2$, it is straightforward to see that iterated elimination of weakly dominated strategies, starting with those cooperating throughout the game, leads to a Nash equilibrium with only defectors.

For the first strategy set, $\Gamma_1$, the Nash equilibrium of $(S_0,S_0)$ is strict since any player deviating would score less. For the second strategy set, $\Gamma_2$, this Nash equilibrium is no longer strict as one of the players could switch to a Follower strategy, still defecting and scoring the same. For the first strategy set, the NE is unique, but for the second one that is not necessarily the case. Since backward induction still applies in the second set, we know that any NE is characterized by defection only, which can be represented by a pair $(S_0,DC_k)$. This is a NE only if the $S_0$ player cannot gain by switching to $CC_{k-1}$ (or to $CC_2$ if $k=2$). This translates into $k P > S + (k-2)R + T$ (for $k>2$) or $T < k P - (k-2)$ for our parameter space, while for $k=2$, we have $2P>S+R$ or $P>1/2$. Furthermore, if $(S_0, DC_k)$ is a NE, then also $(DC_j, DC_k)$, with $j\le k$, is a NE. 

This means that in a part of the payoff parameter region, for $P<1/2$, there is only one NE, while for larger $P$ values there are several NEs, with increasing number the closer to $1$ the parameter $P$ is. Note, though, that any additional NE here is characterized by defection from the first round, corresponding to the unique subgame perfect Nash equilibrium $(S_0, S_0)$. This illustrates the fact that in the NE one can switch from a pure defector to a Follower without reducing the payoff. If this happens under genetic drift in evolutionary dynamics, the situation may change so that Convincers may benefit and cooperation can emerge.

\section{Dynamic Behaviour and Stability Analysis}
\label{numerical_study}

The dynamic behaviour and the stability properties of the fixed points are investigated both analytically and numerically, for the two strategy sets presented in Section \ref{subsect_selectstrat}. In each case, we investigate the dependence of the payoff parameters, $T$ and $P$, as well as the mutation rate and game length, $\epsilon$ and $N$, respectively. We are mainly interested in which influence the presence of Convincers and Followers has on the stability of fixed points and its impact on the dynamics. This will be illustrated in three different ways as follows. First, we present and examine a few specific examples of the evolutionary dynamics and discuss the qualitative difference between the strategy sets. For the simple strategy set, we show analytically that the fixed point associated with the Nash equilibrium remains stable for a sufficiently small mutation rate. A numerical investigation is then performed for the extended strategy set, characterising the fixed point stability. Finally, for different lengths of the game, we study realisations of the evolutionary dynamics from an initial condition of full cooperation ($x_{N} = 1$ and $x_i=0$ for $i \neq N$, with $x_k$ denoting the fraction of strategy $S_k$ in the population) to examine to which degree cooperation survives and whether the dynamics is attracted to a fixed point or characterized by oscillations. By studying variation of the dynamics over the different regions in the parameter space and changing the payoff parameters $T$ and $P$ (by adhering to the inequalities between $T,P,R,S$ as described in Section \ref{sect_evoldynamics}), it is studied how the population evolves in various games over time. 

\subsection{Dynamics with the Simple and the Extended Strategy Set}
\label{exampledynamics}

\begin{figure}
\centerline{
$\begin{array}{cc}
\noindent \includegraphics[width=8cm]{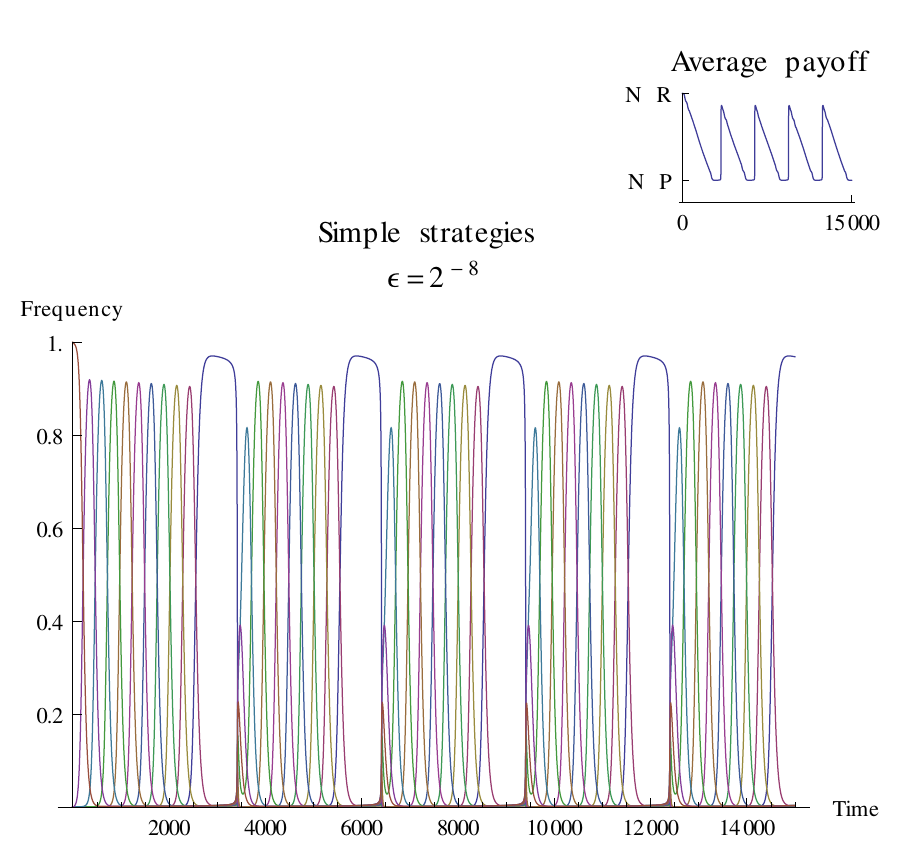} &
\includegraphics[width=8cm]{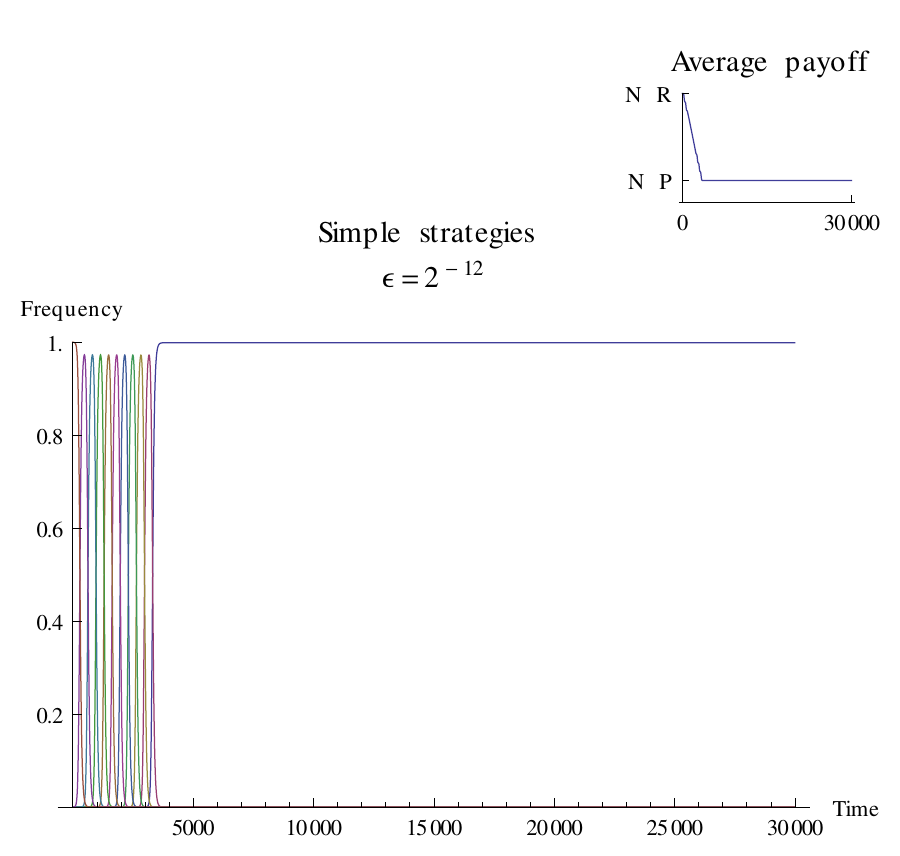} \\ 
\includegraphics[width=8cm]{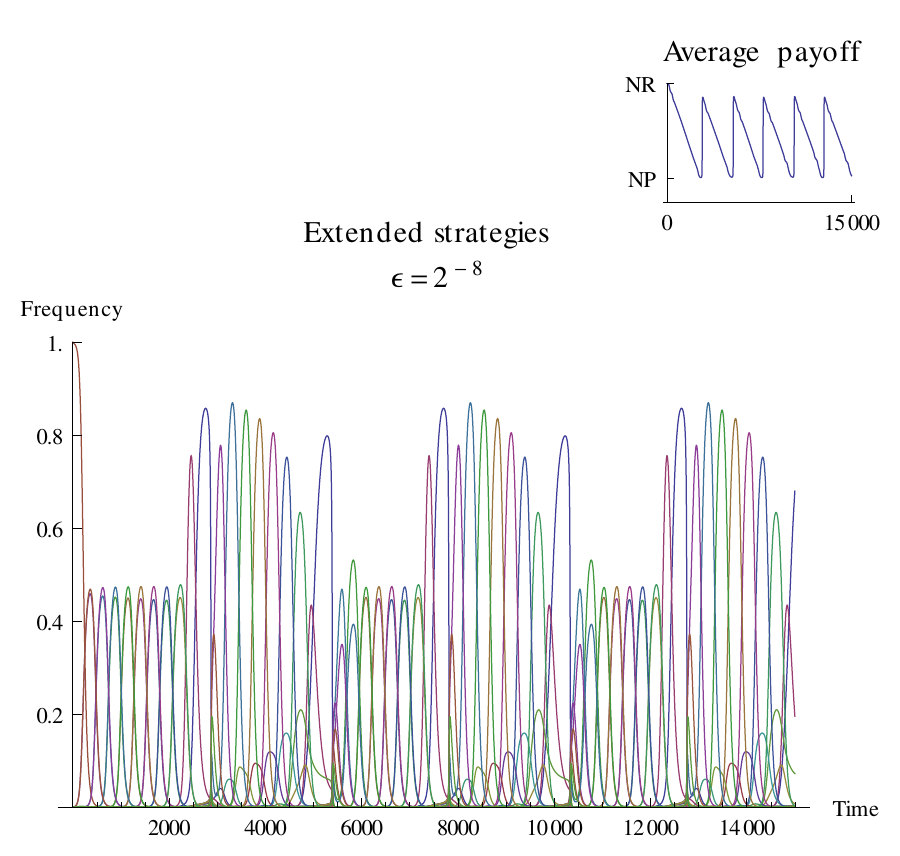} &
\includegraphics[width=8cm]{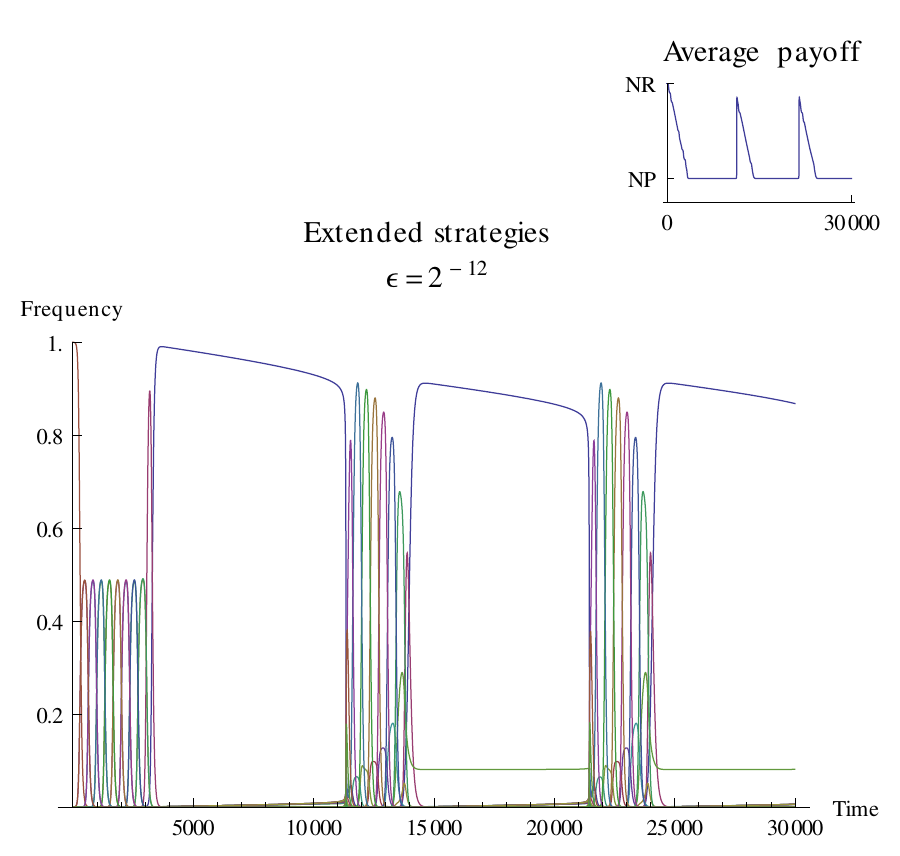}
\end{array}$
}
\caption{\label{fig_example1} Illustration of the dynamics for a particular game (P=0.2, T=1.33) for 10-round Repeated Prisoners' Dilemma ($N=10$). Each main plot displays how the frequency of different strategies changes in the population as a function of time. Smaller subplots show how the population average payoff changes with time, with $NR$ and $NP$ denoting full cooperation and defection, respectively. Top: simple strategy set ($S_0,...,S_{10}$). Below: the extended strategy set, which includes also Convincers ($CC_2,...,CC_{10}$) and Followers ($DC_2,...,DC_{10}$). When lowering $\epsilon$, the extended strategy set exhibits meta-stability with recurring cooperation, while for the simple strategy set cooperation disappears.}
\end{figure}

Realisations of the evolutionary dynamics, Equation~(\ref{eqnMutatorReplicator}), for both the simple and the extended strategy sets are shown in Figure~\ref{fig_example1}. For particular payoff values and a 10-round game, the mutation rate $\epsilon$ is varied to illustrate how it affects the dynamics.

\newpage First, we consider the case with the simple strategies ($S_0,...,S_{10}$) in Figure~\ref{fig_example1}. From an initial state of full cooperation, with a population consisting only of fully cooperative $S_{10}$ players, the dynamics will, for both levels of mutation rate, lead to a gradual unraveling of cooperation to a point where $S_0$, full defection, dominates the population. The first step of this unraveling occurs because $S_9$ defecting in the final round will have higher payoff than $S_{10}$. At this stage $S_0$ is much worse off, but the population goes through a series of transitions which reminds of a backward induction process. This can also be seen in terms of average payoff, as illustrated in Figure~\ref{fig_example1}. When the population is in this non-cooperative mode a positive mutation-rate $\epsilon$ may offset the situation. For the higher mutation rate, cooperative strategies re-emerge after a period of influx from other strategies. The mutations gradually introduce cooperative behaviour to a critical point where some degree of cooperation has a selective advantage over full defection, and we see a shift in the level of cooperation. After a while, cooperative behaviour is again overtaken by full defection and a cyclic behaviour becomes apparent. Comparing with the next realisation of the simple strategies we see that this can happen only when the mutation rate is high enough. As mutation rate becomes smaller, here illustrated with $\epsilon=2^{-12}$, there is no re-appearance of cooperation. When the mutation rate gets too low, strategies other than defection are kept on a level that is too low to promote further cooperation. This demonstrates that the mutation rate can affect whether cooperation re-appears or not.

Second, we consider the case with extended strategies (the $3N-1$ strategies $S_0,...S_{10}$, $CC_2,...,CC_{10}$, $DC_2,...,DC_{10}$) shown at the bottom of Figure~\ref{fig_example1}. We observe that adding Convincer and Follower strategies changes the dynamics, but with a similar unraveling of cooperation. For both mutation rates, the trajectories have periods with defection dominating in the population, indicated by average payoff, but the population does not seem to stabilize. Lowering the mutation rate increases the time between the outbreaks of cooperation. Contrary to the simple strategy set, the system does not seem to settle close to full defection. The explanation is due to the Follower strategies being able to gradually enter the population by getting the same payoff as the strategy of full defection. At a critical point, there are sufficiently many Followers, which make mutations to Convincers successful and cooperation re-enters the population for a period.

In the next section, we will investigate the dynamics and the stability characteristics of both the simple and the extended strategy sets in detail, varying the payoff parameters over the full ranges, and investigating the behaviour in the limit of diminishing mutation rate.

\subsection{Existence of Stable Fixed Points}
\label{fixedPointExistence}

We now turn to examine the existence of stable fixed points in the dynamics for low mutation rates. For the simple strategy set the following proposition holds.

\paragraph{Proposition 1:}

For the simple set of strategies, $\Gamma_1$, the fixed point associated with the Nash equilibrium, dominated by strategy $S_0$, is stable under the replicator-mutation dynamics, if the mutation rate is sufficiently small (but positive). 

\paragraph{Proof 1:} See Appendix A.\\

Our results for the extended strategy set, $\Gamma_2$, are based on numerical investigations: by using an eigenvalue analysis of the Jacobian of the replicator-mutation dynamics, Equation~(\ref{eqnMutatorReplicator}), we determine for which parameters $T, P$ and $\epsilon$ the dynamics is characterised by stable fixed points. We are especially interested in the case of lowering $\epsilon$ towards 0 to examine whether fixed points become stable when mutation rate is sufficiently small, like in the case of the simple strategy set.

\begin{figure}
\centerline{
$\begin{array}{cc}
\includegraphics[width=8cm]{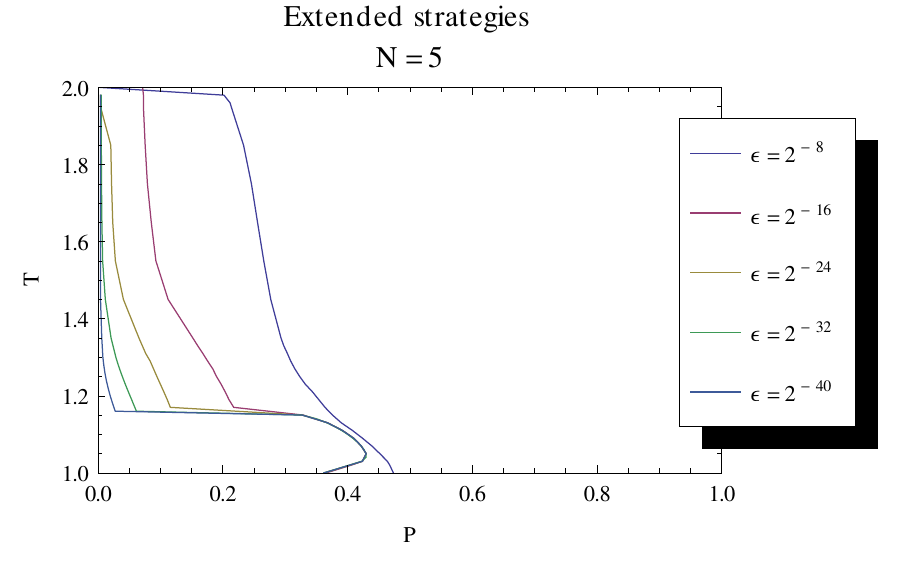} &
\includegraphics[width=8cm]{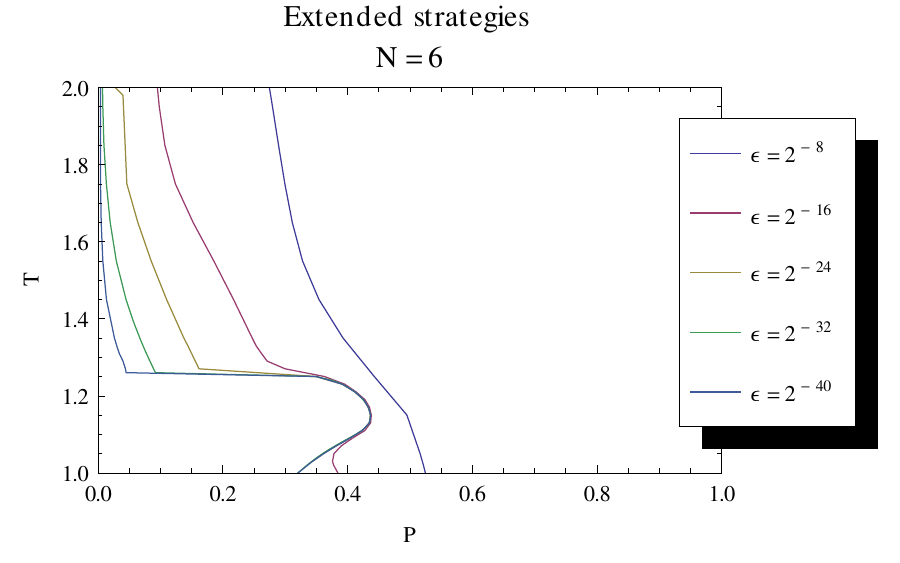} \\
\includegraphics[width=8cm]{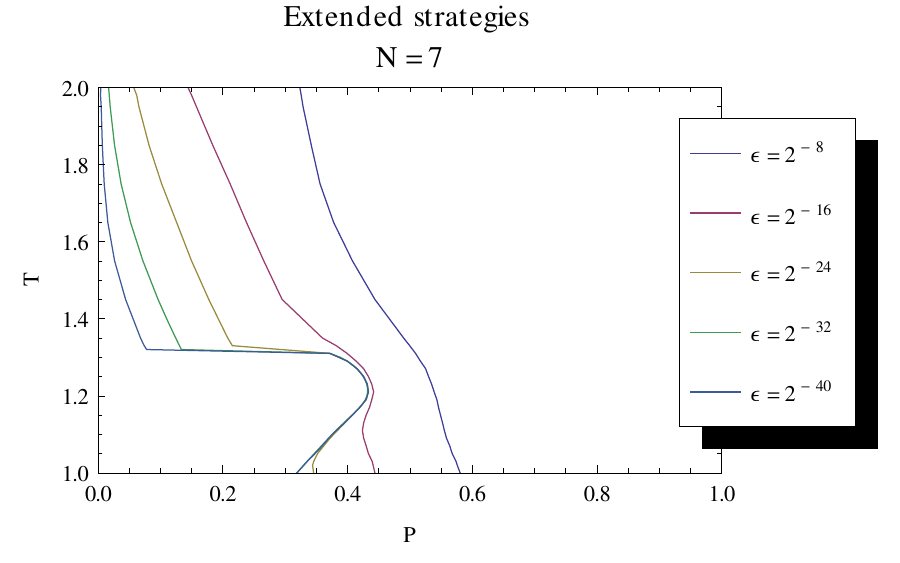} &
\includegraphics[width=8cm]{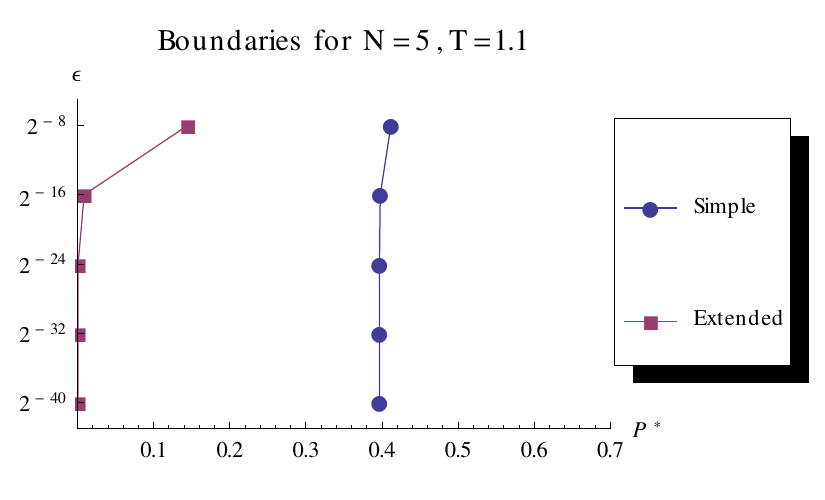} 
\end{array}$
}
\caption{\small \label{fig_fixpoints} Numerical analysis showing which games have stable fixed points. Stable fixed points have been found to the right of a given line, at which they disappear if $P$ is decreased further. Lowering the mutation rate $\epsilon$ turns more evolutionary dynamics into having stable fixed points. Runs for $N=5,6,7$ indicate a convergence, as the mutation rate is decreased, towards a fraction of games being without stable fixed points, as seen in the lower left of the parameter space for the different game lengths. To the bottom right the boundary is shown for a particular $T=1.1$.
}
\end{figure}

This stability analysis over the parameter space and the results are presented for different lengths of the game in Figure~\ref{fig_fixpoints}. Stable fixed points exist to the \emph{right} of the line corresponding to a specific $\epsilon$, while to the \emph{left} of the line no stable fixed points were found. When decreasing the mutation rate $\epsilon$ towards 0, we see that an increasing fraction of parameter space is characterised by a stable fixed point. Unlike the case for the simple strategy set, the numerical investigation shows a convergence of the delimiting line between stable and unstable fixed points, indicating that there is a remaining region in parameter space (with low $P$ and low $T$) for which fixed points are unstable in the limit of zero mutation rate.

Note from the discussion in Section \ref{sectionIntroduction} that while these results show the existence of stable fixed points, it is left to consider whether the population dynamics would actually converge to such states. Next, we turn to consider population dynamics over the parameter space and its outcome.

\subsection{Recurring Phases of Cooperation}
\label{repeated}

Motivated by the findings above in Section \ref{exampledynamics}, we now turn to study recurrent cooperative behaviour in the evolutionary dynamics. We investigate if and for which games a population, starting from initial cooperation as before, settles into a mode of oscillations that at some point in the cycle brings the system back to a higher level of cooperation \footnote{An instance of the dynamics was counted as oscillating when the average payoff $A$ repeatedly returns to at least $5\%$ above full defection, {\em i.e.} $A>1.05NP$. Frequently, it was the case that the oscillations had phases of cooperation well above the $5\%$ threshold.}. 

The interesting case is when mutation rates are small: higher mutation rates introduce a background of all different strategies, which can be seen as artificially keeping up cooperative behaviour in the population. To avoid this effect, we investigate the dynamics with $0<\epsilon<<1$, and, in particular, numerical analyses were made with a series of decreasing mutation rates. The population is initialized as before, described in Section \ref{exampledynamics}.

\begin{figure}[htbp]
\centerline{
$\begin{array}{cc}
\includegraphics[width=8cm]{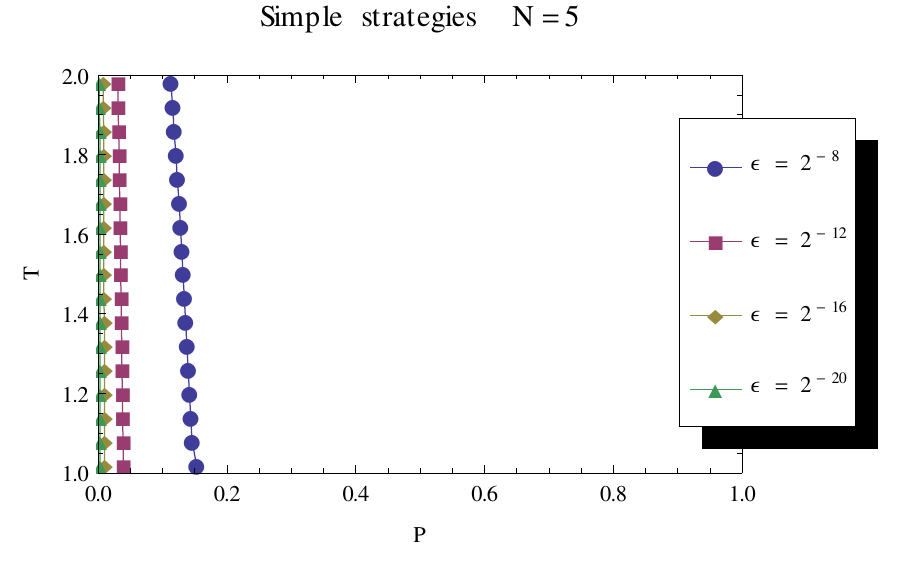} &
\includegraphics[width=8cm]{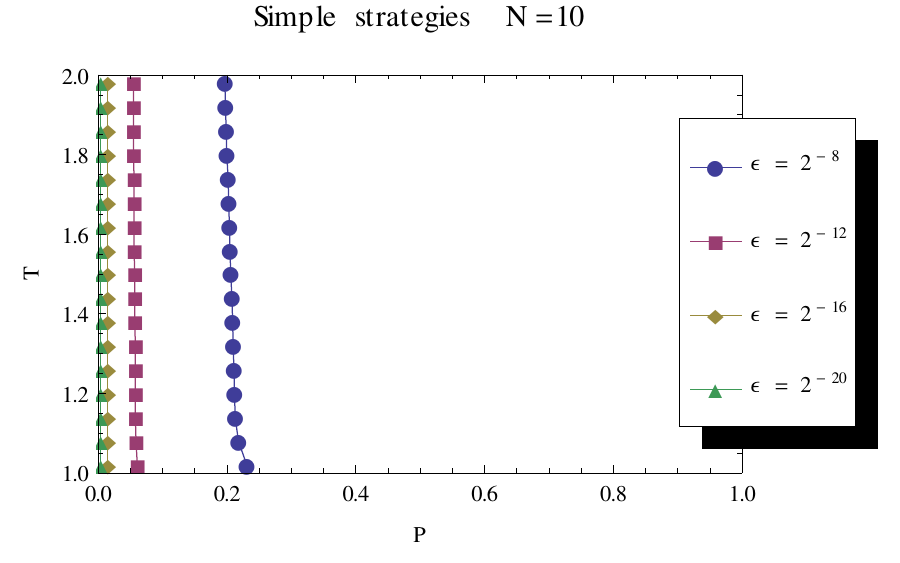} \\
\includegraphics[width=8cm]{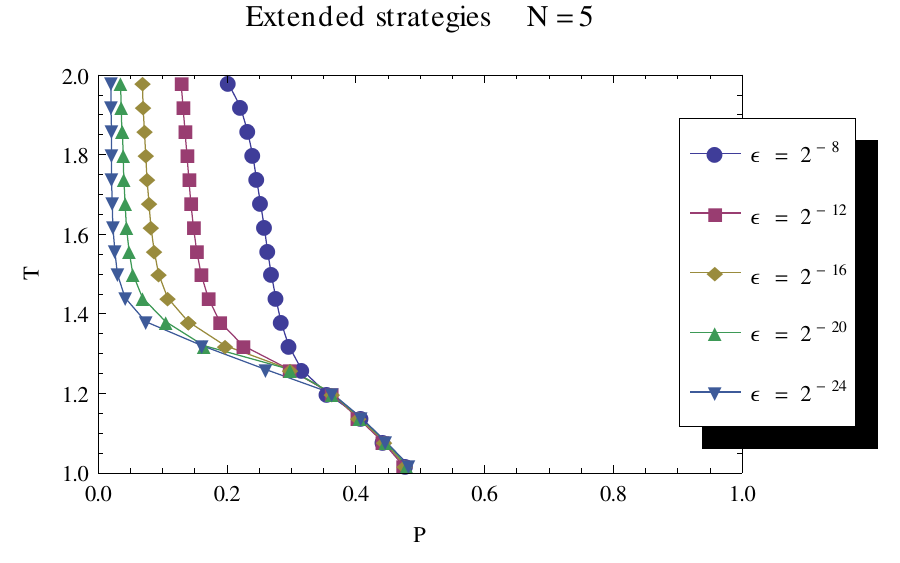} & 
\includegraphics[width=8cm]{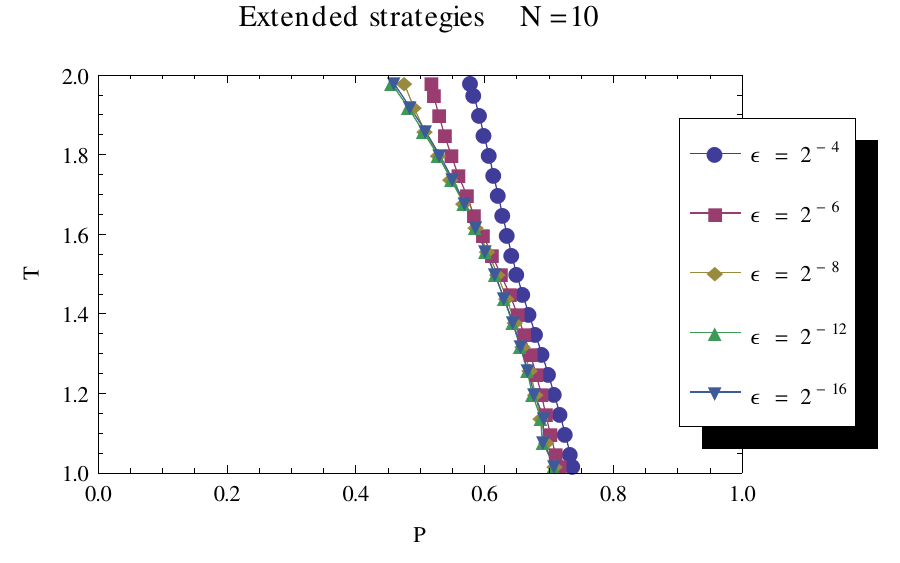} \\
\includegraphics[width=8cm]{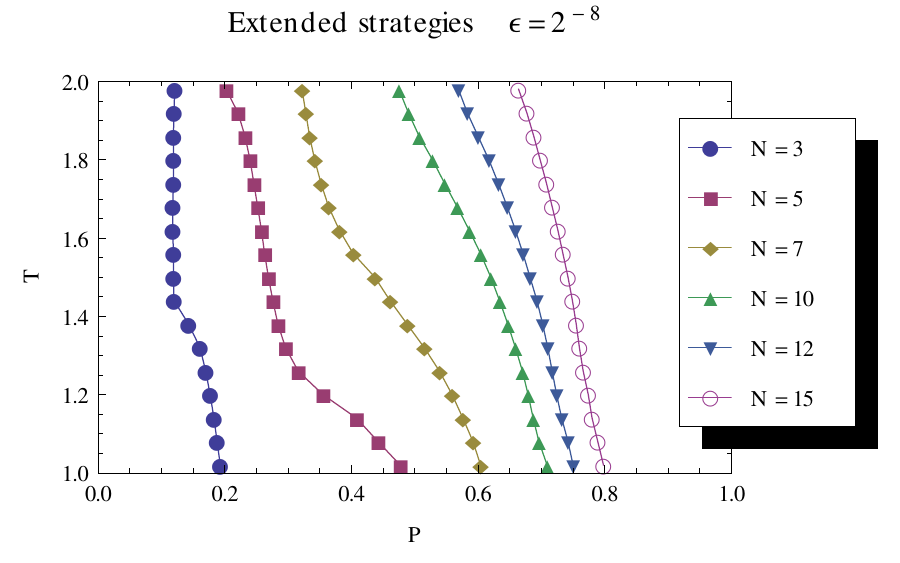} &
\includegraphics[width=8cm]{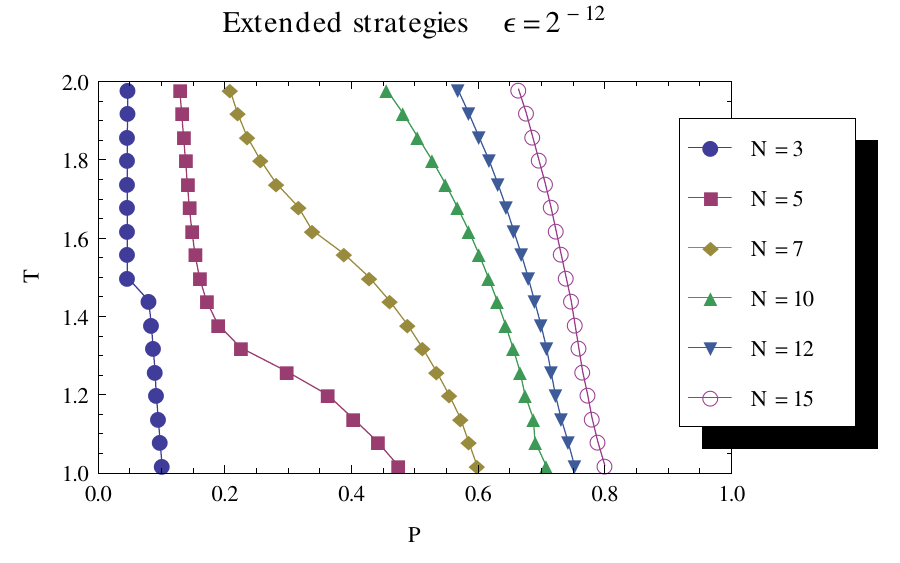} \\
\end{array}$
}
\caption{\label{phaseplots1} Parameter diagram showing which games have recurrent cooperation in the evolutionary dynamics from a starting point of initial cooperation. Top row: simple strategy set. Middle row: extended strategy set with Convincers and Followers. In the graphs, recurrent cooperation exists to the left of the line and a fixed point with defectors characterizes the behaviour to the right. For simple strategies, lowering $\epsilon$ steadily reduces the part of parameter space dominated by recurrent cooperation. In the bottom row, the delimiting line is shown for a variety of game lengths and for two levels of mutation (left and right), illustrating the fact that the longer the game the larger the parameter region supporting cooperative phases in the evolutionary dynamics.
}
\end{figure}

First, we characterise the simple and the extended strategy set for different game lengths, varying mutation rates, and varying the parameters of $T$ and $P$. Figure~\ref{phaseplots1} displays the results for games of 5 and 10 rounds in the top and the middle row. For a given $T$ a line denotes the border for which games to the \emph{left} have recurring phases of cooperation, while games to the \emph{right} are characterised by a fixed point dominated by defection. For the simple strategy set, we observe that by decreasing mutation rate, the fraction of games where cooperation recurs becomes smaller and disappear. This suggests that as we make the mutation rate sufficiently small, cooperation will die out in the case of simple strategies, as is expected from Proposition 1 stating that the fixed point dominated by the $S_0$ strategy becomes stable. On the other hand, for the extended strategy set, a considerable fraction of games seem to offer recurring phases of cooperation despite lowering the mutation. For the 10-round game, the line describing the critical parameters is seen to converge as the mutation rate decreases, {\em i.e.}, there is a large part of the parameter region for which the dynamics is not attracted to a fixed point. For the $5$-round game, this occurs at least in part of the parameter space. The bottom graphs in Figure~\ref{phaseplots1} show that the longer the game, the larger is the parameter region for $T$ and $P$ in which the fixed point is avoided and recurring periods of cooperation are sustained. It should be noted, though, that as the mutation rate decreases, the part of the cycle in which there is a significant level of cooperation decreases towards zero. This is due to the slow genetic drift that brings $DC_k$ strategies back into the population, which eventually make it possible for the $CC_k$ strategies to re-establish a significant level of cooperation.

\subsection{Co-existence between Fixed Point Existence and Recurring Cooperation}

The discussion in Section \ref{fixedPointExistence} left the question of whether stable fixed points are reached by population dynamics, and we found that it is not necessarily so in Section \ref{repeated}. This was illustrated for the extended strategy set for low $\epsilon$. Combining the different findings by considering their boundaries in the parameter space, this suggests an additional property of the evolutionary dynamics. By considering the joint results of our findings (in Figures~\ref{fig_fixpoints} and \ref{phaseplots1}), one can note that there is a part of parameter space for which there is co-existence between a stable fixed point of defect strategies and a stable cycle with recurring cooperation. 

\newpage
\section{Discussion and Conclusion}
\label{discussion_conclusion}

A key point in game theory is that a player's strategic choice must consider the strategic choice the opponent is making. For finitely repeated games, backward induction as a solution concept has become established by assuming player beliefs being based on common knowledge of rationality. However, this assumption says nothing about how players would react to a deviation from full defection---a deviation from the Nash equilibrium---since it a priori rules out actions and reactions that exemplify other ways of reasoning. 

Motivated by the general importance of backward induction and what has been called its "paradox", we have introduced an evolutionary analysis of the interaction in a population of strategies that react differently to out-of-equilibrium play in the first round of the game. We have shown how extending a strategy set for this possibility, in the special case of the repeated Prisoners' Dilemma, allows for stable limit cycles in which cooperative players return after a period of defection. The introduction of Convincers and Followers, representing both strategies that try to establish cooperation and strategies that are capable of responding to that, are made in a way to preserve the structure of the selected strategy set so that elimination of weakly dominated strategies leads to full defection.

For the simple strategy set, as the mutation rate becomes sufficiently small, the cyclic behaviour disappears and the system is attracted to a stable fixed point. The stability of this fixed point was shown analytically for a sufficiently small mutation rate $\epsilon$.

For the extended strategy set, for low levels of mutation, the numerical investigation of fixed point stability and oscillatory modes indicates that, for a certain part of payoff parameter space, the evolutionary dynamics does not reach a stable fixed point but stays in an oscillatory mode, unlike the case of simple strategy set. We characterise our results by a detailed quantitative analysis of where this occurs: showing how the length of the repeated game and the mutation rate affects the boundaries of this region.

One of the main results of the study is an affirmative answer to the question whether different responses to out-of-equilibrium play in the first round can make the dynamics avoid fixed points, and the corresponding Nash equilibrium. Additionally, the fixed point analysis showed the co-existence of a stable fixed point and stable oscillations with recurring phases of cooperation. This means that a system with different responses to out-of-equilibrium play may be found far from its possible stable fixed point. Taken together, this illustrates that the Nash equilibrium play can be unstable at the population level when mutations make explorations off the equilibrium path possible.

This paper contributes to the backward induction discussion in game theory, but more broadly to the study of repeated social and economic interaction. Many models, typically much larger and less transparent ones, of social and economic systems involve agents. If solving these systems means finding the Nash equilibria, then one may doubt whether that is a good representation of rational behaviour except under certain conditions as we have discussed in the paper. We have shown that strategies corresponding to the Nash equilibrium cannot be taken for granted when they interact and compete with strategies that act and respond differently to out-of-equilibrium play.

\newpage
\section*{Acknowledgements}

Financial support from the Swedish Energy Agency is gratefully acknowledged. We would also like to thank two anonymous reviewers for constructive criticism and for inspiring us to prove Proposition 1. We also thank David Bryngelsson for valuable comments on the introduction.

\begin{appendix}
\section*{Appendix A. Stability of fixed point in the simple strategy set for small mutation rates}

In order to show that the Nash equilibrium fixed point at zero mutation rate, $\epsilon = 0$, continuously translates into a stable fixed point as the mutation rate becomes positive, $\epsilon > 0$, we investigate the fixed point more thoroughly. First, we note that the stability of the fixed point without mutations ($\epsilon=0$), characterised by $x_1=1$ (strategy $S_0$ dominating), can be determined by the largest eigenvalue of the Jacobian matrix ($\partial \dot{x}_i / \partial x_j$) derived from the dynamics, Equation~(\ref{eqnMutatorReplicator}), where we use the notation $\dot{x}_i = dx_i/dt$. One finds that the largest eigenvalue is given by $\lambda_\text{max}= - P < 0$, which shows that the fixed point is stable. This is also known from the stability analysis of the finitely iterated game by Cressman \cite{Cressman1996Evolutionary}. 

We will now proceed with reformulating the fixed point condition, for positive mutation rate $\epsilon>0$, which is a set of $n$ polynomial equations given by $\dot{x}_i = 0$ (with $i=1,...,n$), into an equation of only one variable $x_1$. Based on this we show that the Nash equilibrium fixed point for zero mutation rate $\epsilon=0$, characterised by $x_1=1$, continuously moves into the interval $x_1 \in [0,1[$, with retained stability.

We let index $i$ denote strategy $S_{i-1}$ in the simple strategy set ($i=1,...,n$), where the number of strategies is $n=N+1$. Due to the structure of the repeated game for the simple strategy set, determining $u(i,j)$, we can establish pair-wise relations between $x_k$ and $x_{k+1}$ for $k=1,2,...,n-1$, at any fixed point. We use the slightly more compact notation $s_{i,j}=u(i,j)$, for the score for a strategy $i$ against $j$.

For $x_1$ and $x_2$ (strategies $S_0$ and $S_1$), using Equation~(\ref{eqnMutatorReplicator}), we have
\begin{align}
\frac{d x_1}{d t} = x_1 \left( s_{1,1}x_1 + s_{1,2}(1-x_1) - \bar{s} - \epsilon \right) + \frac{\epsilon}{n} = 0
\end{align}
\begin{align}
\frac{d x_2}{d t} = x_2 \left( s_{2,1}x_1 + s_{2,2}x_2 + s_{2,3}(1-x_1-x_2) - \bar{s} - \epsilon \right) + \frac{\epsilon}{n} = 0 \;
\end{align}
where we have used the fact that $s_{1,2}=s_{1,j}$ (for $j>2$), and $s_{2,3}=s_{2,j}$ (for $j>3$). At a fixed point, for $\epsilon>0$, all strategies are present, $x_i>0$, and we can divide these equations by $x_1$ and $x_2$, respectively. Then, taking the difference between the equations gives us an equation for the relation between $x_1$ and $x_2$,
\begin{align}
\frac{\epsilon}{n} \left( \frac{1}{x_1}-\frac{1}{x_2} \right) + x_1 + (T-P) x_2 - (1-P) = 0 \;
\end{align}
where we have used that $s_{2,3}-s_{1,2}=1-P$, $s_{1,1}-s_{2,1}-s_{1,2}+s_{2,3}=1$, and $s_{2,3}-s_{2,2}=T-P$. Solving this quadratic equation gives $x_2$ as a function of $x_1$,
\begin{align}
x_2=f_A(x_1)
\end{align}
with the function $f_A(x)$ defined by
\begin{align}
\label{eqn_fA}
f_A(x) = \frac {1} {2 (T - P)} \left (x - (1 - P) + \frac {\epsilon'} {x} \right) \left (
-1 + 
\sqrt {1 + \frac{4 (T-P) \epsilon'}{\left (x - (1 - P) + \frac {\epsilon'} {x} \right)^2}} \; \right) \,
\end{align}
Here we have introduced $\epsilon'=\epsilon / n$. In the same way, for $x_{k-1}$ and $x_k$ (for $k=3, ..., N-1$), the fixed point implies that 
\begin{align}
\frac{d x_{k-1}}{d t} = x_{k-1} \left( \sum_{j=1}^{k-2} s_{k-1,j}x_j + s_{k-1,k-1}x_{k-1}
+ s_{k-1,k}\left(1-\sum_{j=1}^{k-1} x_j \right) - \bar{s} - \epsilon \right) + \frac{\epsilon}{n} = 0
\end{align}
\begin{align}
\frac{d x_k}{d t} = x_k \left( \sum_{j=1}^{k-2} s_{k,j} x_j + s_{k,k-1}x_{k-1}+ s_{k,k}x_{k}
+ s_{k,k+1}\left(1-\sum_{j=1}^{k} x_j \right) - \bar{s} - \epsilon \right) + \frac{\epsilon}{n} = 0 \;
\end{align}
where we have used the fact that $s_{k,j}=s_{k,k+1}$ for $j>k$. Again, the difference between these equations results in a relation between $x_k$ and all $x_j$ with $j<k$ (for $k=3, ..., N-1$). Since $s_{k-1,j}=s_{k,j}$ for $j<k-1$, the difference can be written
\begin{align}
\frac{\epsilon}{n} \left( \frac{1}{x_{k-1}}-\frac{1}{x_k} \right) + P x_{k-1} + (T-P) x_k - (1-P)\left( 1-\sum_{j=1}^{k-1} x_j \right) = 0 \;
\end{align}
where the score differences, $s_{k-1,k-1}-s_{k,k-1}= P-S=P$, $s_{k,k+1}-s_{k,k}= T-P$ and, $s_{k-1,k}-s_{k,k+1}= P-R=P-1$, have been used. This results in an expression for $x_k$ in terms of all $x_j$ (with $j<k$), 
\begin{align}
x_{k} = f_B\left(x_{k-1},1-\sum_{j=1}^{k-1} x_j\right) 
\end{align}
with the function $f_B(x,w)$ defined by
\begin{align}
\label{eqn_fB}
f_B(x,w) &=
\frac {1} {2 (T - P)} \left (P x - (1 - P) w + \frac {\epsilon'} {x} \right) \left (
- 1 + 
\sqrt {1 + \frac{4 (T-P) \epsilon'}{\left (P x - (1 - P) w + \frac {\epsilon'} {x} \right)^2}} \; \right) \,
\end{align}
Finally, using the same approach for the last two variables, $x_{n-1}$ and $x_{n}$, the equation for the relation between these can be written
\begin{align}
\frac{\epsilon}{n} \left( \frac{1}{x_{n-1}}-\frac{1}{x_n} \right) + P x_{n-1} + (T - 1) x_n = 0 \;
\end{align}
The last variable can then be expressed as
\begin{align}
x_{n} = f_C(x_{n-1}) 
\end{align}
with the function $f_C(x)$ defined by
\begin{align}
\label{eqn_fC}
f_C(x) = 
\frac{1}{2(T-1)}\left(P x+\frac{\epsilon' }{x}\right) \left(-1+
\sqrt{1 + \frac{4 (T-1) \epsilon'}{\left (Px + \frac {\epsilon'} {x} \right)^2}} \; \right) \,
\end{align}
This means that we can recursively express the fixed point abundancies $x_k$ of strategies $k=2,...,n$ in terms of $x_1$, using Eqs.~(\ref{eqn_fA}), (\ref{eqn_fB}), and (\ref{eqn_fC}). Together with the normalisation constraint, this results in an equation with only one variable, $x_1$, that determines the fixed points. In other words: Summation over all $x_k$ and subtraction of 1 gives us a function of $x_1$, $F(x_1)$, with zeroes at the fixed points, 
\begin{align}
F(x_1) = \sum_{k=1}^{n} x_k - 1 \;
\end{align}
Note that we can extend the function $F$ also to the case $\epsilon=0$, since all functions $f_A, f_B$ and $f_C$ are identical to zero in that case and we simply get $F(x_1)=x_1-1$, capturing the Nash equilibrium fixed point $x_1=1$. (Note, though, that in this extension none of the other fixed points at $\epsilon=0$ are captured; any $x_k=1$ defines a fixed point for zero mutation rate, but all except the first one are of no interest for us.)

All functions $f_A, f_B$ and $f_C$ are continuous and bounded, since $g(y)=y(1-\sqrt{1+1/y^2})$ is. We also see that $g(y)\rightarrow 0$ as $|y| \rightarrow \infty$. Thus, $F(x)$ being composed of these functions is continuous in the interval $[0,1]$. 

\begin{figure}
\center
\includegraphics[width=10cm]{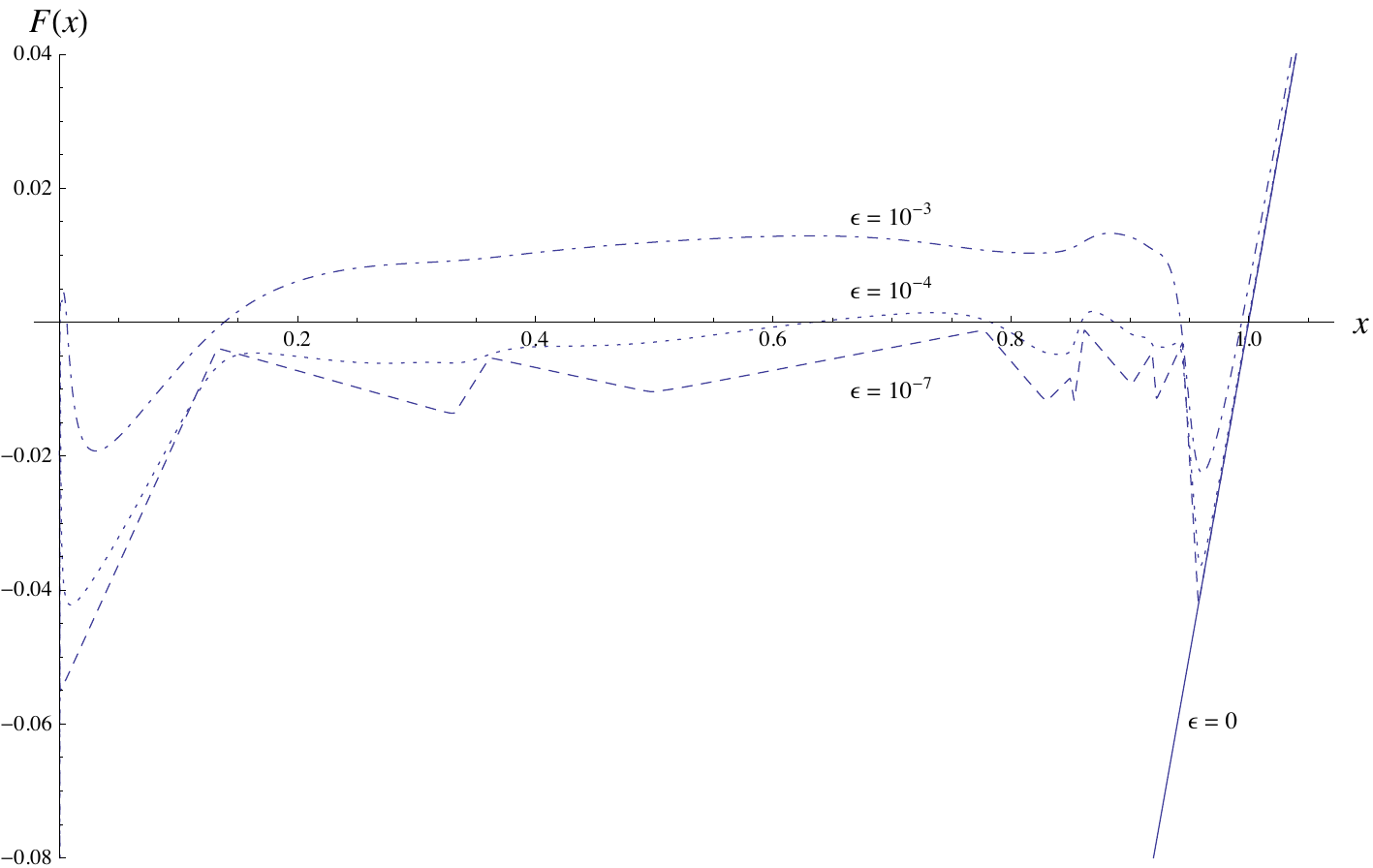}
\caption{\small \label{fig_F} The function $F(x)$ for the 5-round PD game with $T=1.05, P=0.15$, and four different mutation rates $\epsilon$.}
\end{figure}

As an example, Figure~\ref{fig_F} illustrates $F(x)$ for $n=N+1=6, T=1.05, P=0.15$ with four different choices of the mutation rate $\epsilon$. The graph shows that as the mutation rate increases the fixed point at $x_1=1$ moves to the left and also that new fixed points may emerge. Most importantly, though, the graph indicates a discontinuous change of the function for $x$ below about $0.95$, when going from $\epsilon=0$ to $\epsilon>0$. In order to show that the fixed point always moves continuously from $x_1=1$, we need to be sure that any such a discontinuity is bounded away from $x_1=1$.

We already know that for $\epsilon=0$ there is a zero in $x=(1,0,...,0)$. We want to show that as $\epsilon$ increases, this fixed point, characterised by $F(x_1)=0$ for $x_1=1$, continuously moves into the unit interval, $0<x_1<1$. We accomplish this by performing a series expansion of $F(x)$ around $x=1$ and $\epsilon=0$, by showing that $F'(x_1)=1$ for $x_1=1$, that $F$ is increasing with $\epsilon$ in the neighbourhood of $x_1=1$, and that the coordinates $x_k$ ($k>1$) of the fixed point move continuously with $x_1$ when sufficiently close to 1.

The first part is straightforward: At zero mutation rate, $\epsilon=0$, we have already noted that $F(x_1)=x_1-1$, and thus the derivative $F'(x_1)=1$.

Next, we need to show that, at least sufficiently close to the fixed point and for sufficiently small $\epsilon$, $F$ increases with $\epsilon$. We do this term by term in the sum of $F$. First assume that the point $x_1$ is close to fixed point value $1$ at zero mutation rate, {\em i.e.}, $x_1=1-\delta$, and that $\epsilon$ is small, so that $\epsilon << P$ and $\delta << P$. The first term in $F$ is $x_1$ and does not depend on $\epsilon$. The second term, $x_1$, is given by $f_A$, 
\begin{align}
x_2 &= f_A(1-\delta) = 
\frac {1} {2 (T - P)} \left (P-\delta + \frac {\epsilon'} {1-\delta} \right) \left (
-1 + \sqrt {1 + \frac{4 (T-P) \epsilon'}{\left (P-\delta + \frac {\epsilon'} {1-\delta} \right)^2}} \; \right) = \nonumber \\
& = \frac{\epsilon'}{P} \left( 1 + \frac{\delta}{P} + O(\epsilon') \right) + O(\epsilon')^2+ O(\delta)^2 \;
\end{align}
This term thus increases with $\epsilon'$. For the next term $x_3$, using Equation~(\ref{eqn_fB}), we find that to first order in $\epsilon'$ and $\delta$,
\begin{align}
x_3 &= f_B(x_2, 1-x_1-x_2) = f_B\left(\frac{\epsilon'}{P}, \delta-\frac{\epsilon'}{P} \left( 1 + \frac{\delta}{P} \right)\right) = \nonumber \\ 
& = \frac{\epsilon'}{P} \left(1 + \frac{2-P}{P}\delta+ O(\delta)^2 \right) + O(\epsilon')^2 + O(\delta)^2 \;
\end{align}
More generally, for $2<k<n$, repeatedly using Equation~(\ref{eqn_fB}), we find 
\begin{align}
x_k &= \frac{\epsilon'}{P} \left(1 + \frac{k-P(k-1)}{P}\delta+ O(\delta)^2 \right) + O(\epsilon')^2 + O(\delta)^2 \;
\end{align}
and finally, using Equation~(\ref{eqn_fC}), that 
\begin{align}
x_{n} &= \frac{\epsilon'}{P} \left(1 + \frac{(N-1)-P(N-2)}{P}\delta+ O(\delta)^2 \right) + O(\epsilon')^2 + O(\delta)^2 \;
\end{align}
From the linearisation we can conclude that when $x_1$ is sufficiently close to 1, {\em i.e.}, $\delta << P$, the change in $F(x_1)$ from an increase of $\epsilon'$ is given by the $n-1$ terms $x_2, ..., x_n$, or $dF(x_1)/d\epsilon' \sim (n-1)/P > 0$. Adding the fixed point constraint, $F(x_1)=0$, to the linearisation determines $x_1$ and thus $\delta$ in terms of $\epsilon'$. To first order in $\epsilon$ the fixed point is given by
\begin{align}
x_1 &\sim 1 - (n-1) \frac{\epsilon'}{P} + O(\epsilon)^2 = 1-(n-1) \frac{\epsilon}{n P} + O(\epsilon)^2 \;, \text{and} \\
x_k &\sim \frac{\epsilon}{n P} + O(\epsilon)^2 \; \text{ for } k=2,..., n\;
\end{align}
where we have switched back to the original $\epsilon$. This analysis shows that as $\epsilon$ increases from 0, the fixed point gradually moves into the unit interval. 

Since the fixed point is changed continuously as the mutation rate increases from $\epsilon=0$, the eigenvalues of the Jacobian also change continuously. The fixed point at $\epsilon=0$ is stable, with the largest eigenvalue being $\lambda_\text{max}=-P<0$, and we can conclude that for sufficiently small $\epsilon>0$, the real part of the largest eigenvalue remains negative.
From this we can conclude that \emph{the fixed point associated with the Nash equilibrium in the finitely repeated Prisoner's Dilemma remains stable in the simple strategy set if the mutation rate is sufficiently small}. This concludes the proof of Proposition 1.

\end{appendix}

\bibliographystyle{ieeetr}
\makeatletter
\renewcommand\@biblabel[1]{#1. }
\makeatother
\bibliography{games-22020-proofreading-MinorCorrections-ARXIV}
\noindent \\
\noindent This article is an open access article distributed under the terms and conditions of the Creative Commons Attribution license
(http://creativecommons.org/licenses/by/3.0/).

\end{document}